%% file: main.tex
\documentclass[amsfonts, amssymb, amsmath, twocolumn, showkeys, nofootinbib,superscriptaddress, twoside, aps, prx]{revtex4-2}
\usepackage[english]{babel}
\usepackage[utf8]{inputenc}
\usepackage[colorinlistoftodos, color=green!40, prependcaption]{todonotes}
\usepackage{tabularx}
\usepackage{amsthm}
\usepackage{mathtools}
\usepackage{physics}
\usepackage{xcolor, color}
\usepackage{graphicx}
\usepackage{adjustbox}
\usepackage{placeins}
\usepackage[T1]{fontenc}
\usepackage{lipsum}
\usepackage{csquotes}
\usepackage{svg}
\usepackage[pdftex, pdftitle={Article}, pdfauthor={Author}]{hyperref} 
\usepackage{soul}
\usepackage{multirow}

\begin{document}

\title{Turquoise Magic Wavelength of the ${}^{87}$Sr Clock Transition}
    
\author{G. Kestler}
    \affiliation{Department of Physics, University of California San Diego, California 92093, USA}

\author{R. J. Sedlik}
    \affiliation{Department of Physics, University of California San Diego, California 92093, USA}

\author{E. C. Trapp}
    \affiliation{Department of Physics, Harvard University, Cambridge, MA 02138, USA}

\author{M. S. Safronova}
    \affiliation{Department of Physics and Astronomy, University of Delaware, Newark, Delaware 19716, USA}
    
\author{J. T. Barreiro}
    \affiliation{Department of Physics, University of California San Diego, California 92093, USA}
    
\begin{abstract}

Optical lattice clocks of fermionic strontium offer a versatile platform for probing fundamental physics and developing quantum technologies. The bivalent electronic structure of strontium gives rise to a doubly-forbidden atomic transition that is accessible due to hyperfine mixing in fermionic strontium-87, thus resulting in a sub-millihertz natural linewidth. Currently, the most accurate optical lattice clocks operate on this narrow transition by tightly trapping strontium-87 atoms in a {\em magic} optical lattice at 813~nm. {\em Magic} wavelengths occur where the Stark shifts of both the ground and excited states are equivalent, thus eliminating any position and intensity-dependent broadening of the corresponding transition.  Theoretical calculations of the electronic structure of strontium-87 have also predicted another {\em magic} wavelength of the clock transition at 497.01(57)~nm. In this work, we experimentally measure the novel {\em magic} wavelength to be $497.4363(3)$~nm. Compared to the 813~nm {\em magic} wavelength, 497~nm is closer to the strong 461~nm dipolar transition of strontium, resulting in larger atomic polarizability by an order of magnitude, providing deeper traps with less optical power. The proximity to the 461~transition also leads to an enhanced sensitivity of 334(10)~Hz/(nm\,$E_{R}$) at the {\em magic} wavelength. 

\end{abstract}

\maketitle

\section{Introduction}

Cold atom experiments enable key insights into fundamental physics and provide a platform for quantum science and technologies. Gases of neutral atoms are laser-cooled to micro- and nano-kelvin temperatures and optically trapped in engineered light fields for matter-wave interferometers~\cite{Wang2005, Moan2019, Nemirovsky2023}, neutral atom-based quantum computers~\cite{Saffman2016, Weiss2017}, and precision atomic clocks~\cite{Takamoto2005, Boyd2007, Bloom2014, Aeppli2024}. Each platform is developed around the electronic properties of an atomic species; in particular, bivalent alkaline-earth atoms offer a unique toolbox.

Among the alkaline-earth atoms, strontium is particularly well-suited for emulating many-body physics and applications in quantum science and technology. With two valence electrons, strontium's electronic structure offers broad ($\approx$30~MHz) and narrow ($\approx$7~kHz) linewidth atomic transitions for laser cooling, which are utilized to produce atomic clouds at temperatures near 1~$\mu$K with only a two-stage magneto-optical trap. Loading these cold atoms into optical lattices and engineered light fields requires less power and shallower traps than other atomic species~\cite{Kestler2022}. With the bosonic isotope, ${}^{88}$Sr, the 7.4~kHz intercombination line is used for sensitive matterwave interferometry~\cite{Aguila2018} as well as in high-resolution spectroscopy for determining light shifts~\cite{Kestler2019, Kestler2022} and probing surface interactions, such as the Casimir-Polder effect~\cite{Martin2017, Cook2017}. Arrays of atomic clocks have also been implemented using the doubly forbidden sub-millihertz `clock' transition of strontium-88 ~\cite{Covey2019}. Precise optical control of the fermionic isotope, ${}^{87}$Sr, which has a nuclear spin of 9/2, realized fermionic spin-momentum lattices for many-body physics ~\cite{Lauria2022}, inspired quantum computers with fermionic hardware~\cite{GonzalezCuadra2023}, and enabled precision spectroscopy for the most accurate atomic clocks to date~\cite{Aeppli2024}.

Atomic clocks of ${}^{87}$Sr rely on recoil-free spectroscopy through the tight confinement of the cold atoms in a deep optical lattice. The bivalent electronic structure of strontium provides a spherically symmetric ground state with no angular momentum, $J=0$. The ${}^{1}S_{0}-{}^{3}P_{0}$ transition, with $J=0 \rightarrow J'=0$, is thus doubly-forbidden by both the spin and angular momentum selection rules~\cite{Trautmann2023}. However, this transition is optically accessible by hyperfine mixing in the fermionic isotopes~\cite{Boyd2007} or magnetic-field-induced mixing in the bosonic isotopes~\cite{Taichenachev2006}. High-resolution Rabi spectroscopy is performed by exciting the lattice-trapped atoms into the long-lived ${}^{3}P_{0}$ state before repumping them back to the ${}^{1}S_{0}$ ground state. The highly sensitive spectroscopic platform has measured fermionic interactions~\cite{Goban2018}, gravitational redshift within millimeter scales~\cite{Bothwell2022}, and black-body radiation shifts~\cite{Aeppli2024}.

Optical lattices generate an energy potential for the atoms through an AC Stark shift, where the atoms experience an induced dipole moment from the time-varying trapping field. The direction and strength of the energy shift depend on the trapping field intensity and the atomic polarizability, which in turn depend on the wavelength and the electronic energy state of the atoms. If the ground state, ${}^{1}S_{0}$, and the excited state, ${}^{3}P_{0}$, of the `clock' transition experience different energy shifts, the transition resonance will depend on %
the field intensity at the atom position, leading to an undesired broadening. Employing a {\em magic} wavelength for the optical lattice ensures both the ground and excited state experience the same AC Stark shift, thus preserving the narrow `clock' transition. Optical traps at 813.428~nm have been used for {\em magic} `clock' lattices but have small atomic polarizability compared to other transitions and thus require more optical power for deep lattice traps. 

We experimentally confirm a predicted {\em magic} wavelength of strontium's `clock' transition near 497~nm, benchmarking the accuracy of the atomic theory. The polarizability of the $^3P_0$ state near 497~nm is dominated by the contribution of the $5s5d$~$^3D_1$-$5s5p$~$^3P_0$
 $5p2$~$^3P_1$-$5s5p$~$^3P_0$ matrix elements. Knowledge of these matrix elements is crucial for further refining the BBR shift estimates in the Sr clock at room temperature.   The excellent agreement between the theoretical and experimental values demonstrates the validity of the theoretical values for these matrix elements and uncertainty estimates.

When compared to the standard 813.428~nm optical lattice clocks, this novel {\em magic} wavelength is only slightly detuned from the strong 461~nm transition, yielding an order of magnitude improvement in atomic polarizability for both the ground and excited state of the clock transition. We can thus realize much deeper traps with the same optical powers and waists as the current optical lattice clocks.

\section{Theory}

To compute the Sr magic wavelength, we calculate the $^1S_0$ and $^3P_0$ polarizabilities using an approach that combines configuration interaction (CI) and coupled cluster methods. We refer to this method as  CI+all-order approach \cite{SafKozJoh09,SafPorSaf13,HeiPar20,pci}.
In this method, the energies and wave functions are determined from the time-independent multiparticle Schr\"odinger equation
\begin{equation}
H_{\rm eff}(E_n) \Phi_n = E_n \Phi_n,
\label{Heff}
\end{equation}
where the effective Hamiltonian is defined as
\begin{equation}
H_{\rm eff}(E) = H_{\rm FC} + \Sigma(E),
\end{equation}
where $H_{\rm FC}$ is the Hamiltonian in the frozen core approximation and $\Sigma$ is the energy-dependent correction, which
takes into account virtual core excitations computed using the linearized coupled-cluster method \cite{SafKozJoh09,pci}.
The scalar dynamic polarizability $\alpha(\omega)$ separates into three parts:
\begin{equation}
\alpha(\omega) = \alpha_v(\omega) + \alpha_c(\omega) + \alpha_{vc}(\omega),
\label{alpha}
\end{equation}
where $\alpha_v$ is the valence polarizability, $\alpha_c$ is the ionic core polarizability. A small term $\alpha_{vc}$
is included due to the presence of two valence electrons to restore the Pauli principle by slightly modifying the core
polarizability. The $\alpha_{c}$ and $\alpha_{vc}$ terms were evaluated in the random-phase approximation (RPA).

The valence part of the polarizability is determined by
solving the inhomogeneous equation of perturbation theory in the
valence space. It is approximated as
\begin{equation}
(E_v - H_{\textrm{eff}})|\Psi(v,M^{\prime})\rangle = D_{\mathrm{eff},q} |\Psi_0(v,J,M)\rangle
\label{eq1}
\end{equation}
for a state  $v$ with the total angular momentum $J$ and projection
$M$ \cite{kozlov99a}.  The $D_{\mathrm{eff}}$ is the effective dipole operator that includes RPA corrections. 
This approach accounts for all intermediate, high-lying discrete states and the continuum, and provides the total theory value for the valence scalar polarizability at a given frequency.
 
 We also repeat all computations with $\Sigma$ constructed using the second-order many-body perturbation theory to assess the uncertainties (CI+MBPT approach), as this provides an estimate of omitted correlation corrections for cases where no experimental matrix elements are available.

To improve the accuracy of the computations, we recalculate all the dominant contributions to all polarizabilities (for each wavelength point) using the experimental energies \cite{NIST} and recommended values of the electric dipole matrix elements, where available, using
the formula \cite{MitSafCla10}:
\begin{equation}
    \alpha_{0v}=\frac{2}{3(2J+1)}\sum_k\frac{{\left\langle k\left\|D\right\|v\right\rangle}^2(E_k-E)}{(E_k-E)^2-\omega^2}, \label{eq-1} 
\end{equation}
where
$\omega$ is assumed to be at least several linewidths off
resonance with the corresponding transitions and ${\left\langle k\left\|D\right\|v\right\rangle}$ are the reduced electric-dipole matrix elements.

\begin{figure}[t!]
    \includegraphics[width=\columnwidth]{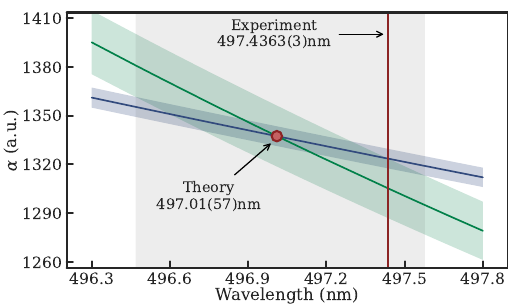}
    \caption{Theoretical prediction of the ${}^{1}S_{0}$ ground state (blue) and ${}^{3}P_{0}$ excited state (green) polarizabilities. The shaded blue (green) represents the theoretical uncertainties $\delta\alpha$ in the ground (excited) state polarizabilities. The {\em magic} wavelength occurs at their overlap (red circle at 497.01(57)~nm), with the uncertainty represented by the grey shaded area. The experimental value (red line at 497.4363(3)~nm), falls within the theoretical prediction.}
  \label{fig:theory}
\end{figure}

We plot the final values of the $^1S_0$ and $^3P_0$  polarizabilities in Fig.~\ref{fig:theory} and determine the magic wavelength to be 497.01(57)~nm. To estimate the uncertainty in the value of the magic wavelength, we also plot $\alpha+\delta \alpha$ and $\alpha-\delta \alpha$ values of the $^1S_0$ and $^3P_0$ polarizabilities. 
The uncertainty is determined as the largest difference of the $\alpha+\delta \alpha$ and $\alpha-\delta \alpha$  crossings with the central crossing that determines the magic wavelength.

\section{Experimental Setup}

Cold atom experiments are typically performed in cycles that start from a hot gas of atoms and progress through several stages of laser cooling, culminating in a destructive measurement of the atom number. The experimental cycle in this work is adapted from previous work with this apparatus~\cite{Lauria2022} that has a standard two-stage magneto-optical trap (MOT) operating first on the ${}^{1}S_{0}-{}^{1}P_{1}$ transition at $460.86$~nm. The second stage MOT uses the narrow-line ${}^{1}S_{0} (F=9/2)-{}^{3}P_{1} (F=11/2)$ transition at $689.4$~nm and a weaker {\em stirring} beam resonant with the ${}^{1}S_{0} (F=9/2)-{}^{3}P_{1} (F=9/2)$ to mix the $m_{F}$ substates~\cite{Stellmer2014}. A high-NA objective in the downward MOT path produces asymmetric beams in the vertical direction, introducing a small degree of spin polarization and resulting in unequal populations of the plus and minus Zeeman spin-projected states. However, all 10 spin states are still present. We load the $\approx$1~$\mu$K atoms into a crossed dipole trap and perform evaporative cooling for 5.43 seconds to reach a degenerate Fermi gas (DFG) with $T/T_{F}=0.30(1)$.

\begin{figure}[t!]
    \includegraphics{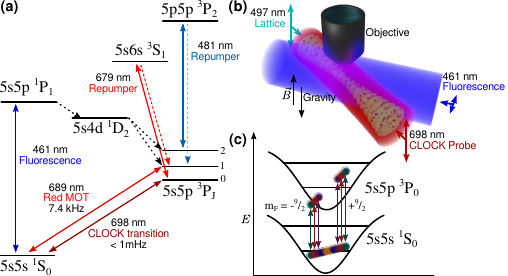}
    \caption{(a) Relevant levels and transitions for ${}^{87}$Sr used in the experiment. (b) Experimental setup of the optical lattice at 497~nm. The polarization of the optical lattice and `clock' probe is vertical and parallel to the direction of gravity. Fluorescence is collected from an objective (NA=0.8) and driven by a horizontally polarized 461-nm probe beam operating on the strong dipolar transition. An $\approx$17-G magnetic field, $\vec{B}$, is applied along the direction of gravity and sets the quantization axis parallel to the probe polarization for spin-resolved spectroscopy. (c) The atoms are adiabatically loaded into the lowest vibrational energy level of the lattice. A resonant laser drives $\pi$-transitions for each $m_{F}=\{\pm 5/2, \pm 7/2, \pm 9/2\}$ sub-level between the $5s^{2}\ {}^{1}S_{0}$ ground state and $5s5p\ {}^{3}P_{0}$ excited state.}
  \label{fig:setup}
\end{figure}

The DFG is loaded adiabatically into a horizontally oriented, retro-reflected optical lattice at $\approx$497~nm. We co-propagate a narrow-linewidth, cavity-locked resonant probe beam with the optical lattice to interrogate the ${}^{1}S_{0}-{}^{3}P_{0}$ doubly-forbidden `clock' transition of the trapped atoms (see Fig.~\ref{fig:setup}(b)). Both the lattice and probe beams are vertically polarized along the direction of gravity to ensure that we drive $\pi$-transitions. For a broad initial determination of the {\em magic} wavelength, we excite the lattice-trapped atoms to the metastable ${}^{3}P_{0}$ `clock' state with saturated probe light ($\Omega = 2\pi\times 21$~Hz) for 400 milliseconds and remove any residual ground state atoms in the lattice by driving the ${}^{1}S_{0}-{}^{1}P_{1}$ transition. A later, more precise measurement is performed with a $\pi$-pulse of 47~ms and a Fourier-limited linewidth of $\approx$20~Hz. The probe laser is locked to an ultra-stable cavity (Stable Laser Systems) with a specified linewidth of 1 Hz. The excited state atoms are then repumped back to the ground state via the $5s6s {}^{3}S_{1}$ and $5p^{2} {}^{3}P_{2}$ states with $\approx$679~nm light and $\approx$481~nm light for 10 milliseconds. Both repumper lasers are scanned over hundreds of GHz at a rate of $\approx$2~kHz to ensure repumping of all 10 Zeeman spin states. After the excited state atoms are repumped, we collect fluorescence on the ${}^{1}S_{0}-{}^{1}P_{1}$ transition through our high NA=0.8 objective onto an EMCCD camera.

When the atoms are optically trapped in a {\em magic} wavelength lattice, the ${}^{1}S_{0}-{}^{3}P_{0}$ resonance is independent of the trap depth. For multiple wavelengths from 497.2~nm - 497.55~nm, we load atoms into the lattice at various final depths from 20$E_{R}$ to 44$E_{R}$, where $E_{R}=(\hbar k)^{2}/2m$ is the recoil energy associated with a lattice photon. We apply a $\approx$17~G magnetic field along gravity and perform spin-resolved spectroscopy of the $\pm 9/2, \pm 7/2$, and $\pm 5/2$ transitions (see Fig.~\ref{fig:setup}(c)) and average each relative substate to cancel magnetic field and non-linear polarization effects. The linear relationship provides a differential Stark-shift coefficient, which is zero at the {\em magic} wavelength.

\section{Results}

\begin{figure}[t!]
    \includegraphics{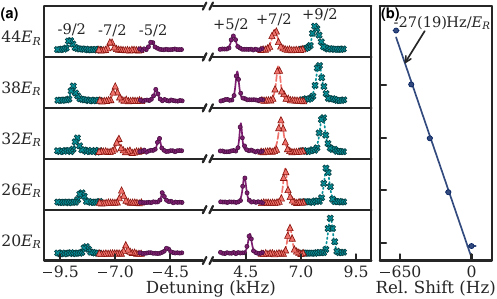}
    \caption{Measurement of the differential Stark shift at 497.35~nm. (a) Narrow line spectroscopy of the $m_{F}=\{\pm 5/2, \pm 7/2, \pm 9/2\}$ spin states. Asymmetric vertical MOT beams produce a slightly spin-polarized sample before loading into the lattice, resulting in uneven excited-state populations of the $\pm$ spin states. (b) Linear relationship of the resonant shift to the lattice trap depth. The differential Stark shift is given by the inverse of the slope in Hz/$E_{R}$.}
  \label{fig:single}
\end{figure}

Figure~\ref{fig:single}(a) illustrates the spectroscopy results at an off-{\em magic} trapping wavelength of 497.35~nm. The $\pi$-polarized probe light ensures we are driving the ${}^{1}S_{0} (m_{F} = \pm5/2, \pm7/2, \pm9/2) - {}^{3}P_{0} (m_{F}=\pm5/2, \pm7/2, \pm9/2$ transitions (see Fig.~\ref{fig:setup}(c)). Different peak heights between the +5/2, +7/2, and +9/2 are proportional to the Clebsch-Gordon coefficients, while the smaller populations of the -5/2, -7/2, and -9/2 Zeeman substates are a result of the asymmetric MOT beams, which give rise to a slight spin polarization of the atomic cloud.

We cancel any magnetic field noise by averaging the resonance of alternating $m_{F}$ states. The quantization axis at the atoms is set along the direction of gravity by a 17.03(5)-G magnetic field generated by the MOT coils in a Helmholtz configuration. The $\approx$109~Hz/G~\cite{Campbell2017} differential level shift between the ${}^{1}S_{0}$ and ${}^{3}P_{1}$ gives rise to a 1.856(5)~kHz separation between neighboring Zeeman substates. Additionally, the spectral lines are broadened to $\approx$180~Hz, a result of the saturated probe power and the AC Stark shift. The latter occurs since the state-dependent energy shifts are different depending on the location of the atoms in the Gaussian radial profile of the lattice. By varying the total trapping depth, a linear relationship between Stark shift and trap depth can be extracted as the differential Stark shift coefficient (DSSC) for a single wavelength (see Fig.~\ref{fig:single}(b)).

A possible systematic error in a single DSSC measurement from Fig.~\ref{fig:single} may occur due to a slow drift in the ultra-stable cavity, which cascades to the laser frequency and thus changes the relative spectroscopy results over time. Since the data point in Fig.~\ref{fig:single} takes around 4 hours to complete, we reduce the drift rate to $<$5~mHz/s and actively correct for the drift during data collection (see Appendix). Density-dependent shifts can also contribute to systematic errors since the atomic density in a lattice site varies by $\approx$25\% across the range of trap depths. However, these shifts are typically on the order of 1~Hz~\cite{Rey2009}, an order of magnitude less than the fit uncertainty, and do not affect a single DSSC measurement.

\begin{figure}[t!]
    \includegraphics{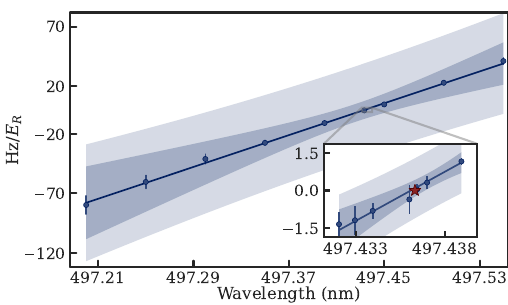}
    \caption{Differential Stark shifts at various wavelengths. The linear relationship provides a sensitivity of 334(10)~Hz/(nm\,$E_{R}$) and the {\em magic} wavelength is given at 497.434(2)~nm. For the inset, we spin-polarize to the $m_{F}=+9/2$ state and decrease the linewidth to 20~Hz, measuring a {\em magic} wavelength of 497.4363(3)~nm indicated by the red star. In both plots, the light shaded regions are the 95\% confidence interval and the dark shaded regions are the 95\% prediction interval.}
  \label{fig:magic}
\end{figure}

\begin{figure}[t!]
    \includegraphics{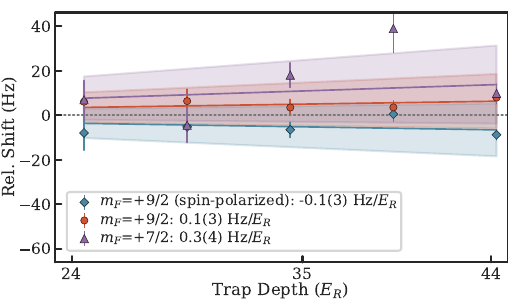}
    \caption{Demonstration of zero shift for a wide range of trap depths from 24~$E_{R}$ to 44~$E_{R}$ confirming the {\em magic} wavelength at 497.4363(3)~nm. The shaded areas represent the uncertainty in the slope, indicating that the zero-slope line falls within this uncertainty.}
  \label{fig:zero_slope}
\end{figure}

We further perform spin-polarization to the $m_{F}=+9/2$ substate and run Fourier-limited spectroscopy using a $\pi$-pulse of 47~ms, providing an $\approx$20~Hz linewidth. We repeat the previous measurement, with only a single Zeeman state, and perform three measurements per power per wavelength. This allows us to determine the {\em magic} wavelength more precisely at 497.4363(3)~nm with a sensitivity of 334(10)~Hz/(nm\,$E_{R}$). We further demonstrate zero shift, within uncertainty, for trap depths in the range of 24~$E_{R}$ to 44~$E_{R}$ (see Fig.~\ref{fig:zero_slope}). Confirming that there is no discrepancy from interaction effects, we also demonstrate a zero shift without spin-polarization and with the $m_{F}=+7/2$ substate.

\section{Outlook}

In this work, we report on a novel {\em magic} wavelength of the ultranarrow strontium `clock' transition at 497.4363(3)~nm. Efforts to produce more accurate clocks rely on close collaboration between theoretical calculations and experimental techniques to reduce the observed linewidth and eliminate shifts and broadening effects~\cite{Bloom2014, Aeppli2024}. Measurement of the novel {\em magic} wavelength offers insight into theoretical calculations of dipole matrix elements for ${}^{87}$Sr.

Observing the `clock' transition in a {\em magic} optical lattice at 497~nm marks an experimental departure from the prevalent 813~nm optical traps with a number of possible avenues to pursue, including arrays of atomic clocks~\cite{Covey2019, GonzalezCuadra2023}, super- and sub-radiant phenomena~\cite{Norcia2016}, and deeply degenerate Fermi gases in optical tweezers~\cite{Viverit2001}. Specifically, for arrays of atomic clocks implemented with tightly focused optical tweezers, the smallest waists achievable with high-NA objectives are diffraction-limited and proportional in size to the wavelength of light. For 497-nm light, the smallest spot sizes are 497/813$\approx$0.6 times smaller. Since optical trap depths are proportional to $\alpha/\omega_{0}^{2}$, one can expect a 30$\times$ improvement in trapping efficiency for equivalent setups. Additionally, the smaller spot size allows for 3$\times$ more traps in the same area, leading to a more scalable and efficient platform for optical clock arrays of single fermions, which can be used to implement fermionic quantum computers~\cite{GonzalezCuadra2023}. The tightly focused optical tweezers also provide a small trapping volume, which can adiabatically load ultracold atoms from the crossed dipole trap. The transfer increases the Fermi energy while keeping the atom temperature relatively constant, thus compressing the Fermi gas and enhancing the degeneracy parameter of the cloud~\cite{Viverit2001}. In terms of optical lattice clocks, the separation between neighboring sites in a 497~nm lattice is $\approx$248.5~nm. With a transition wavelength of 698~nm, this falls in a regime to study super- and sub-radiant phenomena with optical clocks, even more so than the $\approx$400~nm spacing in 813~nm optical lattices~\cite{Norcia2016}. 

\begin{acknowledgements}
 The Office of Naval Research supported the theoretical work (Grant No. N000142512105), while the experimental work was supported by the National Science Foundation Awards No. 1752630 and 2412662.
\end{acknowledgements}

\input{appendices}

\bibliography{references}

\end{document}

%% file: appendices.tex
\begin{appendix}

\section{Spectroscopy Measurement}

\subsection{Broad Scans}
Since each experimental run is $\approx$10 seconds, a single data point in Fig.~\ref{fig:magic} takes $\approx$4 hours. Long-term frequency drift in the laser frequency can cause a systematic error in the spectroscopy measurements. We measure a drift in the ultrastable cavity on the order of 3~kHz per day or 23~mHz/s and implement feed-forward stability by adjusting the laser frequency into the cavity with an electro-optical modulator updated every second. Feed-forward correction risks accidentally narrowing the transition during high-resolution spectroscopy scans, so we partially implement the correction amount to ensure the drift rate is $<5$~mHz/s. The spectroscopic shift at each trap depth is varied out-of-order measuring 44$E_{R}$, 32$E_{R}$, 20$E_{R}$, 26$E_{R}$, and 38$E_{R}$ followed by a second reference at 44$E_{R}$. Any residual drift between the first and last 44$E_{R}$ spectroscopy measurement is accounted for, and the errors are propagated to each data point prior to fitting the line. We also account for magnetic field noise by alternating $m_{F}$-substate scans and averaging both to determine the appropriate Stark shift at the given lattice depth.

Each data point in Fig.~\ref{fig:single} is a single experimental run. Although we do average the alternating $m_{F}$ states and all three substates, we determine the 95\% confidence and prediction intervals using the $n=1$ two-sided Student-t score.

\subsection{Narrow Scans}
For the narrower scans used for the inset of Fig.~\ref{fig:magic}, we spin-polarize to the $m_{F}=+9/2$ substate and measure three cycles through trap depths of 44$E_{R}$, 32$E_{R}$, 20$E_{R}$, 26$E_{R}$, and 38$E_{R}$. Each individual trap depth then has three data points to determine the cavity drift correction, which are averaged for a more precise {\em magic} measurement of 497.4363(3)~nm. The 95\% confidence and prediction intervals utilize the $n=3$ two-sided Student-t score.

\section{Lattice Calibration}

\begin{figure}[b]
    \includegraphics{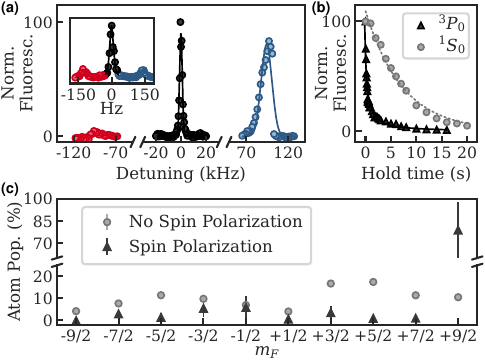}
    \caption{497~nm lattice characterization. (a) Axial and radial (inset) sideband spectroscopy. (b) Lifetime measurements for the ${}^{1}S_{0}$ ground state and ${}^{3}P_{0}$ excited state. The rapid decay of the excited state is due to density-dependent $p$-wave scattering. The density-independent lifetimes of the ground and excited states are 7.2(4) and 8(2) seconds, respectively. (c) Atom populations with and without spin polarization. The asymmetric 689 MOT beams create a slight spin polarization in the positive $m_{F}$ states.}
  \label{fig:lattice}
\end{figure}

We calibrate the lattice by performing sideband spectroscopy for both the axial and radial sidebands (see Fig.~\ref{fig:lattice}(a)) and determining the trap depth as $U_{0}/E_{R} = \nu_{z}^{2}m^{2}\lambda_{T}^{4}/h^{2}$ and $\nu_{z} = \sqrt{32\alpha P/4\pi w_{0}^2 \lambda_{T}^{2} c \epsilon_{0} m}$ where the waist is determined by $w_{0} = \lambda_{T}\nu_{z}/\sqrt{2}\pi\nu_{r}$. Thus, the trap depth is linear with total power, $P$, and determining the axial and radial trap frequencies for a fixed power is enough to characterize the trap depth at all the powers. The relative heights of the blue and red axial sidebands further confirm that we are indeed loading into the ground state of the lattice.

We also measure the trapping lifetimes for atoms in the ground state and excited states (see Fig.~\ref{fig:lattice}(b)). The ground state lifetime is 8(2) seconds, but the excited state lifetime suffers a rapid drop off within the first 500~ms. This decay agrees well with the two-body loss model below~\cite{Bishof2011} rather than an exponential decay, similar to that of the ground state.
$$\dot{n}_{e} = -\Gamma n_{e} - K_{ee}n_{e}^{2}$$
Fitting to this model for the number of atoms over time, $n_{e}(t)$, we extract a density-independent lifetime, $\Gamma$, of 7.2(4) seconds, but accurately determining the two-body loss coefficient, $K_{ee}$, is beyond the scope of this paper. Reducing the atom density and spin-polarizing the trapped atoms recovers an exponential decay for the lifetime.

\end{appendix}